\documentclass[11pt,aps,showpacs]{revtex4}

\usepackage{epsfig}

\newcommand{\nc}{\newcommand}
\nc{\be}{\begin{equation}}
\nc{\ee}{\end{equation}}
\nc{\bea}{\begin{eqnarray}}
\nc{\eea}{\end{eqnarray}}
\nc{\lsim}{\mbox{\raisebox{-.6ex}{~$\stackrel{<}{\sim}$~}}}
\nc{\gsim}{\mbox{\raisebox{-.6ex}{~$\stackrel{>}{\sim}$~}}}
\nc{\gtwid}{\mathrel{\raise.3ex\hbox{$>$\kern-.75em\lower1ex\hbox{$\sim$}}}}
\nc{\ltwid}{\mathrel{\raise.3ex\hbox{$<$\kern-.75em\lower1ex\hbox{$\sim$}}}}

\begin{document}

\vskip -0.2in

\rightline{CERN-TH/2003-172, HD-THEP-03-38, UFIFT-HEP-03-20}

\vskip 0.2in

\title{Dynamics of super-horizon photons during inflation
       with vacuum polarization}

\author{Tomislav Prokopec}
\email{T.Prokopec@thphys.uni-heidelberg.de,
      Tomislav.Prokopec@cern.ch
}
\affiliation{Institut f\"ur Theoretische Physik, Heidelberg
University,
    Philosophenweg 16, D-69120 Heidelberg, Germany}

\affiliation{Theory Division, CERN, CH-1211 Geneva 23, Switzerland}

\author{Richard Woodard}
\email{woodard@phys.ufl.edu}
\affiliation{Department of Physics, University of Florida,
Gainesville, Florida 32611, USA}

\date{\today}

\begin{abstract}

We study asymptotic dynamics of photons propagating in the polarized vacuum
of a locally de Sitter Universe. The origin of the vacuum polarization is
fluctuations of a massless, minimally coupled, scalar, which we model 
by the one-loop vacuum polarization tensor of scalar electrodynamics. 
We show that late time dynamics of the electric field on superhorizon scales
approaches that of an Airy oscillator. The magnetic field amplitude,
on the other hand, asymptotically approaches a nonvanishing constant
(plus an exponentially small oscillatory component), which is suppressed
with respect to the initial (vacuum) amplitude. This implies that 
the asymptotic photon dynamics is more intricate than that of a massive 
photon obeying the local Proca equation.

\end{abstract}


\maketitle

\section{Introduction}
\label{Introduction}

Even though we have recently witnessed significant progress in understanding 
the dynamics of gauge fields (photons) during inflation, our understanding 
remains quite rudimentary. In earlier work~\cite{ProkopecWoodard:2003,
ProkopecTornkvistWoodard:2002a,ProkopecTornkvistWoodard:2002b} we computed
the one loop vacuum polarization from charged, massless, minimally coupled
scalars in a locally de Sitter background. We also evaluated the integral
of the position-space vacuum polarization tensor against a tree order photon 
wave function and found a nonzero result which grows with the same number of 
scale factors as the mass term of the Proca equation. 

What we have so far done has a direct analogue in flat space quantum field
theory. Because the background possesses spacetime translation invariance
it is much simpler to work in momentum space. Gauge and Lorentz invariance 
imply that the vacuum polarization is proportional to the transverse 
projection operator, $\Pi^{\mu\nu}(p) = (p^2 \eta^{\mu\nu} - p^{\mu} p^{\nu}) 
\Pi(p^2)$. Here $\eta^{\mu\nu}$ is the Minkowski metric,
\begin{eqnarray}
  \eta_{\mu\nu} = {\rm diag}(-1,1,1,1) \, , \qquad \eta^{\mu\nu} = 
{\rm diag}(-1,1,1,1) \, ,\label{metric:eta}
\end{eqnarray}
and $p^2 \equiv p^{\mu} p^{\nu} \eta_{\mu\nu}$. One checks that the photon 
remains massless by evaluating $p^2 \Pi(p^2)$ on the tree order mass shell, 
$p^2 = 0$. A nonzero result --- as in the Schwinger model \cite{Schwinger:1962} 
--- proves that quantum corrections change the tree order mass shell, but it 
does not determine the fully corrected mass shell. For that one must find 
the pole of the full propagator, that is, solve for $p^2$ such that,
\begin{equation}
p^2 \Bigl[1 - \Pi(p^2)\Bigr] = 0 \; . \label{fulleqn}
\end{equation}
What we have done for photons during inflation is the analogue of computing 
$\Pi(p^2)$ at one loop and showing that $p^2 \Pi(p^2)\vert_{p^2 =0} \neq 0$. 
What we do in this paper is the analogue of solving equation (\ref{fulleqn}).

To introduce the problem, we remind the reader that we work in the conformal
coordinate system of a locally de Sitter space-time whose metric and inverse 
are,
\begin{eqnarray}
  g_{\mu\nu}  = a^2 \eta_{\mu\nu} \,,\qquad g^{\mu\nu}  = a^{-2} \eta^{\mu\nu}
\, . \label{metric}
\end{eqnarray}
Here $a$ is the scale factor, 
\begin{equation}
 a(\eta) = -\frac{1}{H\eta} \, , \qquad (\eta <0) \, . \label{scale factor}
\end{equation}
The Lagrangean density of massless minimally coupled scalar electrodynamics
in a general metric reads, 
\begin{eqnarray}
{\cal L}_{\Phi QED}  & = & 
                       -  \frac 14\sqrt{-g} g^{\mu\rho}g^{\nu\sigma}
                                   F_{\mu\nu}F_{\rho\sigma} 
                       - \sqrt{-g}g^{\mu\nu}
                         (\partial_{\mu} - ie A_{\mu}) \phi^* 
                         (\partial_{\nu} + i e A_{\nu} ) \phi  
\,,
\label{PhiQED}
\end{eqnarray}
where $F_{\mu\nu} \equiv \partial_{\mu} A_{\nu} - \partial_{\nu} A_{\mu}$.
Its reduction to the locally de Sitter background~(\ref{metric}--\ref{scale 
factor}) is,
\begin{eqnarray}
{\cal L}_{\Phi QED}  & \rightarrow & 
                  -  \frac 14 \eta^{\mu\rho}\eta^{\nu\sigma}
                                   F_{\mu\nu}F_{\rho\sigma} 
                 -a^2  \eta^{\mu\nu}
                        (\partial_{\mu} - ie A_{\mu}) \phi^* 
                        (\partial_{\nu} + i e A_{\nu} ) \phi
\,,
\label{PhiQED:conformal}
\end{eqnarray}
where we made use of $\sqrt{-g} = a^4$. From this form of the Lagrangean, 
it is obvious that the photon field couples conformally to gravity, with 
the trivial rescaling,
\begin{eqnarray}
  F_{\mu\nu} \rightarrow F_{\mu\nu}\qquad (A_\mu\rightarrow A_\mu)
\,.
\label{conformal-rescaling:photon}
\end{eqnarray}
An important consequence is that tree level photon wave functions take the 
same form in conformal coordinates as they do in flat space, $A_\mu \propto 
\epsilon_\mu {\rm e}^{ik\cdot x}$.

On the other hand, the minimally coupled scalar field $\phi$
does not couple conformally to gravity, which can be easily seen 
upon the conformal rescaling, $\phi \rightarrow a^{-1}\phi$ in
Eq.~(\ref{PhiQED:conformal}). This has as
a consequence the extensively studied particle creation during inflation,
known as superadiabatic amplification. At one loop this conformal symmetry
breaking is communicated to the photon sector by the cubic and quartic terms
in~(\ref{PhiQED:conformal}). The main objective of this work is to study how
this changes photon wave functions.

 Provided it couples minimally to gravity, an obvious candidate for 
the scalar field $\phi$ is the charged component
of the fundamental Higgs scalar, which at low energies
manifests as the longitudinal component of the $W^{\pm}$ boson.
Of course, there may be many more charged scalars of this type
whose masses lie between that of the Higgs field, $m_H\sim 10^2$\,GeV,
and the energy scale of inflation, $H\simeq 10^{13}\,{\rm GeV}$. An important 
class of such particles would be the supersymmetric scalar partners
of the standard model quarks and charged leptons.

In the presence of charged scalar fluctuations the vacuum becomes polarized, 
such that in conformal space times~(\ref{metric}) Maxwell's equations 
generalize to~\cite{ProkopecWoodard:2003},
\begin{equation}
\eta^{\mu\nu} \eta^{\rho\sigma} \partial_{\rho} F_{\sigma\nu}(x) \!
+ \!\!
\int \! d^4x' [\mbox{}^{\mu} \Pi_{\rm ret}^{\nu}]\!(x,x') A_{\nu}(x') \! = \! 0
\, . \label{genMax}
\end{equation}
Here $[\mbox{}^{\mu} \Pi_{\rm ret}^{\nu}]\!(x,x')$ denotes the retarded 
vacuum polarization bi-tensor, which has the general form,
\begin{eqnarray}
[\mbox{}^{\mu} \Pi_{\rm ret}^{\nu}]\!(x,x') &\equiv& 
-  [{}^\mu\! P^\nu] \, \chi_e(x,x') 
 + [{}^\mu\! \bar P^\nu]\, \delta n^2(x,x') \, . \quad \label{genPi}
\end{eqnarray}
The two transverse projectors are defined by, 
\begin{equation}
[{}^\mu\! P^\nu] \equiv
        [\eta^{\mu\nu} \eta^{\rho\sigma} - \eta^{\mu\rho} \eta^{\sigma\nu}]
              \partial^{\prime}_{\rho}\partial_{\sigma}
\,,\qquad
[{}^\mu\! \bar P^\nu] \equiv
   \eta^{\mu i} \eta^{\nu j}     [\delta_{ij}\nabla^\prime\cdot\nabla
                      - \partial^\prime_i\partial_j]
\,,
\label{transverse:projectors}
\end{equation}
with $\partial_{\mu} \equiv \partial/\partial x^{\mu}$, $\partial_{\mu}^{
\prime} \equiv \partial/\partial x^{\prime\mu}$. The coefficient of 
$[{}^\mu\! \bar P^\nu]$ in (\ref{genPi}) is,
\begin{equation}
 \delta n^2(x,x') = \chi_e(x,x') + \frac{\chi_m}{1+\chi_m}(x,x')
\,.
\label{deltan2}
\end{equation}
where $\chi_e(x,x')$ and $\chi_m(x,x')$ denote 
the electric and magnetic susceptibilities of the vacuum.
An important consistency check of equations~(\ref{genMax})--(\ref{genPi})
is obtained by taking a derivative $\partial_\mu$ of Eq.~(\ref{genMax}).
The first term vanishes by the antisymmetry of $F_{\sigma\nu}$, while
the integral term vanishes because of the transverse structure
of the vacuum polarization tensor~(\ref{genPi}),
\begin{equation}
  \partial_\mu [{}^\mu\Pi^\nu] = 0 = \partial_\nu^\prime [{}^\mu\Pi^\nu]
\, . \label{dmu:muPinu}
\end{equation}
This is an immediate consequence of
$ \partial_\mu\; [{}^\mu\! P^\nu] = 0$,
$ \partial_\mu\;[{}^\mu\! \bar P^\nu] = 0$,
$ \partial^\prime_\nu [{}^\mu P^\nu] = 0$,
and $ \partial^\prime_\nu [{}^\mu\bar P^\nu] = 0$.

\medskip

In  a locally de Sitter space-time~(\ref{scale factor}) 
there is a simple relation between ${\ell}(x;x')$, the 
invariant length from $x^{\mu}$ to $x^{\prime \mu}$, and the conformal
coordinate interval ${\Delta x}^2(x;x')$,
\begin{equation}
4 \sin^2\left(\frac12 H \ell(x;x') \right) = a a' H^2 {\Delta x}^2(x;x') 
\equiv y(x;x') \; .
\end{equation}
We refer to $y(x;x')$ as the de Sitter length function. The scalar 
propagator can be expressed in terms of $y \equiv y(x;x')$ and the two scale 
factors,
\begin{eqnarray}
\lefteqn{{i \Delta}(x;x') = \frac{H^{D-2}}{(4 \pi)^{\frac{D}2}} \left\{ 2^{D-4}
\Gamma(D-1) \ln(a a') - \pi {\rm cot}(\pi {\scriptstyle \frac{D}2}) 
\frac{\Gamma(D-1)}{\Gamma(\frac{D}2)} \right.}
\nonumber \\
& & \hspace{3cm} \left. + \sum_{n=1}^{\infty} \biggl[ \frac1{n} \frac{
\Gamma(D-1+n)}{\Gamma(\frac{D}2 + n)} \Bigl(\frac{y}4\Bigr)^n - \frac1{n- 
\frac{D}2} \frac{\Gamma(\frac{D}2 - 1 + n)}{\Gamma(n)} \Bigl(\frac{y}4\Bigr)^{
n - \frac{D}2}\biggr] \right\} \; . 
\label{Delta}
\end{eqnarray}
Note that the homogeneous terms (i.e., $y^n$) and the constant terms have
slightly different proportionality constants from our previous 
expression~\cite{OnemliWoodard:2002,ProkopecTornkvistWoodard:2002b}. 
This has been done to make (\ref{Delta}) valid for regulating 
a spacetime in which $D$ will ultimately be taken to some dimension other 
than four~\footnote{We thank Ewald Puchwein for pointing out the 
improvement.}. In $D=4$, the propagator~(\ref{Delta}) reduces to,
\begin{equation} 
 i \Delta(x;x') \; \stackrel{D\rightarrow 4}{\longrightarrow} \;
     \frac{H^2}{4\pi^2}\bigg\{
            	              \frac{\eta\eta'}{\Delta x^2}
                            - \frac 12 \ln(H^2\Delta x^2) 
                            -\frac 14 + \ln(2) 
        	        \bigg\}
\,.
\label{Delta:4dim}
\end{equation}

In the presence of charged scalar fluctuations described by this propagator,
the photon in a locally de Sitter space-time sees a polarized vacuum. This
effect is characterized by the renormalized vacuum polarization bi-tensor,
the one loop result for which is~\cite{ProkopecTornkvistWoodard:2002a,
ProkopecTornkvistWoodard:2002b}, 
\begin{eqnarray}
i \Bigl[\mbox{}^{\mu}\Pi^{\nu}_{\rm ren}\Bigr](x,x') 
   &=& \frac{\alpha}{32\pi^3} 
      \bigg\{
         -\;  [{}^\mu\!P^\nu]\;
          \biggl[
                 \frac{1}{12}\partial^4\Big(\ln^2(\mu^2 {\Delta x}^2) 
                                         -   2\ln(\mu^2 {\Delta x}^2) \Big)
 \nonumber \\
& & \hspace{2.6cm} +\; i\,{16 \pi^2 \over 3} \ln(a) \delta^4(x - x') 
\nonumber\\
&&\hspace{2.6cm}
      + \; \frac{1}{\eta\eta'}\partial^2\Big(\ln^2({H^\prime}^2 {\Delta x}^2) 
                                          +   2\ln({H^\prime}^2 {\Delta x}^2)
                                         \Big)
           \biggr]
\nonumber \\
& & \hspace{1.4cm} 
        +\;  [{}^\mu\!\bar P^\nu]\;
            \Bigl[ \frac{2}{\eta^2 \eta^{\prime 2}}
                    \Big(\ln^2({H^\prime}^2 {\Delta x}^2) 
                      +  4\ln({H^\prime}^2 {\Delta x}^2)
                    \Big)
            \Bigr] 
      \bigg\} 
\, . \quad 
\label{vacpol}
\end{eqnarray}
Here $\mu$ is the renormalization scale, $\alpha \equiv e^2/4\pi$ is the fine 
structure constant, and $H^\prime \equiv 2^{-1}{\rm e}^{1/4} H$.
The polarization tensor~(\ref{vacpol}) was obtained using dimensional 
regularization with the scalar propagator~(\ref{Delta}). It is fully 
renormalized, which means that it is an integrable function of $x^{\prime
\mu}$ when the derivatives are taken after the integration. It is also gauge 
invariant on account of its transversality on both indices~(\ref{dmu:muPinu}),

The renormalized vacuum polarization (\ref{vacpol}) is properly an in-out
matrix element. To convert it into the retarded vacuum polarization 
$[{}^\mu\Pi_{\rm ret}^\nu]$ of Eq.~(\ref{genMax}) we employ the 
Schwinger-Keldysh formalism as has been described at length
elsewhere~\cite{ProkopecTornkvistWoodard:2002b}. The result takes the
same form as (\ref{genPi}) with electric susceptibility,
\begin{eqnarray}
 \chi_e &=& \chi_e|_{\rm flat} 
         +  \chi_e|_{\rm anomaly} 
         +  \chi_e|_{\rm de\;Sitter}
\nonumber\\
 \chi_e|_{\rm flat} &=& \frac{\alpha}{96\pi^2}\,\partial^4
                        \Big[
                             \Big(
                                  \ln(\mu^2\Delta\tau^2)-1
                             \Big)
                             \theta(\Delta\eta)\theta(\Delta\tau^2)
                        \Big]
\nonumber\\
 \chi_e|_{\rm anomaly}  &=& \frac{\alpha}{6\pi}\,\delta^4(x-x'\,)
\nonumber\\
\chi_e|_{\rm de\;Sitter} &=& \frac{\alpha}{8\pi^2}H^2aa'\,\partial^2
                        \Big[
                             \Big(
                                  \ln({H^\prime}^{\,2}\Delta\tau^2)+1
                             \Big)
                             \theta(\Delta\eta)\theta(\Delta\tau^2)
                        \Big]
\, .
\label{vacuum:pol:retarded:2}
\end{eqnarray}
The change in the index of refraction is,
\begin{eqnarray}
\quad\;
\delta n^2 &=& \frac{\alpha}{4\pi^2}H^4a^2{a'}^2\,
                             \Big(
                                  \ln({H^\prime}^{\,2}\Delta\tau^2)+2
                             \Big)
                             \theta(\Delta\eta)\theta(\Delta\tau^2)
\,.
\label{vacuum:pol:retarded:3}
\end{eqnarray}
In these formulae $a\equiv a(\eta)$, $a'\equiv a(\eta')$, $H^\prime = 2^{-1}{
\rm e}^{1/4} H$, $\Delta \tau^2 = (\eta -\eta')^2 - \| \vec x-\vec x^{\,\prime}
\|^2$, $\Delta \eta = \eta-\eta'$, and $\theta(x) = 1$ for $x>0$, $\theta(x) 
= 0$ for $x<0$. 

The electric susceptibility in~(\ref{vacuum:pol:retarded:2})
can be neatly split into three contributions. 
The first contribution $\chi_e|_{\rm flat}$ is conformally invariant.
When expressed in terms of conformal time, this contribution becomes identical
to the one-loop electric susceptibility 
in Minkowski (flat) space-time. This term contains no scale
factors, and it is renormalized precisely as in flat space. 
The second contribution, $\chi_e|_{\rm anomaly}$, comes from the 
trace anomaly, which arises as a consequence
of imperfect cancellation between the counterterm and 
the local one-loop diagram in dimensional 
regularization~\cite{ProkopecTornkvistWoodard:2002b}.
The anomalous contribution depends only weakly on 
the scale factor $a$. The trace anomaly was first considered
in the context of electrodynamics in conformal space-times
in Ref.~\cite{Dolgov:1981}, while for a discussion  
of scalar electrodynamics and nonabelian gauge theories see
Refs.~\cite{BirrellDavies:1982} 
and~\cite{Dolgov:1993,ProkopecTornkvistWoodard:2002b}.
The third contribution to the electric susceptibility, 
$\chi_e|_{\rm de\;Sitter}$ in~(\ref{vacuum:pol:retarded:2})  
is purely inflationary, completely finite and grows linearly
with the scale factors $a$ and $a'$.
And finally, $\delta n^2$ in~(\ref{vacuum:pol:retarded:3}) 
is the second inflationary contribution, which
grows quadratically with the scale factors $a$ and $a'$.

In section \ref{Photon dynamics with vacuum polarization in inflation}
we argue that only the one loop terms $\chi_{e\vert{\rm de\ Sitter}}$ and
$\delta n^2$ need be kept in equation (\ref{genMax}). We also give the
integro-differential equation which describes a spatial plane wave with 
arbitrary time dependence. (The derivation is reserved for the Appendix.) 
In section \ref{Late-time dynamics of superhorizon photons} we analyze 
the integro-differential equation for superhorizon photons at 
asymptotically late times. We show analytically that while the vector
potential approaches a nonzero constant, its time derivative behaves 
like an Airy oscillator. Numerical analysis is also presented. In
section \ref{Discussion} we translate these results into statements 
about the physical electric and magnetic fields.

\section{Photon dynamics with vacuum polarization in inflation}
\label{Photon dynamics with vacuum polarization in inflation}

We shall now study the dynamics of the photon field by including 
only the genuinely new inflationary terms contributing 
at one-loop to the vacuum polarization
tensor, $\chi_e|_{\rm de\;Sitter}$~(\ref{vacuum:pol:retarded:2}) 
and $\delta n$~(\ref{vacuum:pol:retarded:3}), since 
their contribution to the vacuum polarization grows rapidly
with the scale factor. Indeed,  $\delta n\propto a^2$, 
and the electric susceptibility,
$\chi_e|_{\rm de\; Sitter}\propto a$. From 
$\partial_{\sigma} a = \delta^0_{\sigma} H a^2$ it follows however, 
that this term can also give an $a^2$ in Eq.~(\ref{genMax})
and has to, therefore, be treated on an equal footing as $\delta n^2$. 
Moreover, we expect that, when compared with the one-loop contributions to 
$\chi_e|_{\rm de\;Sitter}$ and $\delta n^2$, the contributions of higher loops
can be neglected, since they are down by more powers of $\alpha$, 
but can give no extra powers of $a$. 

A general solution can be expressed as a linear combination of spatial
plane waves in Coulomb gauge,
\begin{equation}
   A_{\mu}(x) = \epsilon_\mu(\vec k,\eta)\; {\rm e}^{i\vec k\cdot \vec x}
\,,\qquad
\epsilon_i k_i = 0
\,. 
\label{vector:field:plane:wave}
\end{equation}
In the Appendix we show that $\epsilon_0(\vec{k},\eta)$ vanishes in the absence 
of sources, just as it does in flat space. The Appendix also describes how 
to perform the integral over $\vec {x}^{\,\prime}$ to obtain the following 
integro-differential equation for the time dependent polarization vector,
\begin{eqnarray}
&&-(\partial_0^2 + k^2)\epsilon_i(\vec k,\eta)
  + \frac{\alpha H^2}{\pi k}
  \bigg[a(\partial_0^2 + {k}^{\,2})\!\int_{\eta_0}^\eta\! d\eta' a'
           Y_1\Big(k\Delta\eta,\frac{k}{H^\prime}\Big)\,
            \epsilon_i(\vec k,\eta^\prime) 
\nonumber\\
&&\hspace{4.3cm}
   +\;   H^2a^2
       \!\int_{\eta_0}^\eta\! d\eta' {a'}^2
       \sin(k\Delta\eta)\,\epsilon_i(\vec k,\eta^\prime)
  \bigg]
  = 0
\, .  \quad \label{eom:Ai}
\end{eqnarray}
Here $Y_1(x,\zeta)$ is,
\begin{equation}
Y_1(x,\zeta)
      =  \sin(x)\big[2\ln(2x/\zeta)+2
      +   {\rm ci}(2x)-\gamma_E-\ln(2x)\big] 
      -  \cos(x)\big[{\rm si}(2x)+\pi/2\big] 
\,.\quad
\label{Y1}
\end{equation}
Upon acting the time derivatives upon the first integral, 
Eq.~(\ref{eom:Ai}) can be recast as,
\begin{eqnarray}
&&\!\!\!\!\!\!\!\!\!\!\!\!\!\!\!
 (\partial_0^2 + k^2)\epsilon_i(\vec k,\eta)
 + a^2 \,\frac{2\alpha H^2}{\pi}
   \bigg\{\bigg(\ln(a)
             - \frac 54
         \bigg)
            \epsilon_i(\vec k,\eta)
       -  \frac{H^2}{k} \int_{\eta_0}^\eta d\eta' {a'}^2
            \sin(k\Delta\eta)\epsilon_i(\vec k,\eta') 
\nonumber\\
 &&\hspace{3.5cm}
     +\; \frac 1a  \int_{\eta_0}^\eta\! d\eta' a'
            \frac{1-\cos(k\Delta\eta)}{\Delta\eta}
                \,\epsilon_i(\vec k,\eta^\prime)
      - \int_{\eta_0}^\eta\! d\eta'
        \ln\Big(1-\frac{\eta}{\eta'}\Big)
                            \partial_0^{\prime} \epsilon_i(\vec k,\eta^\prime)
\nonumber\\
      &&\hspace{3.5cm}-\; \ln\Big(1-\frac 1a\Big)\epsilon_i(\vec k,\eta_0)
   \bigg\}
 = 0
.
\label{eom:Ai:2}
\end{eqnarray}

This integral-differential equation is one of our main results. It should
describe photon dynamics accurately from the beginning of inflation, when
$a(\eta_0) = 1$ and the one loop terms are negligible, all the way to late
times, when $a \gg 1$ and the one loop terms become important. Another of
our main results is that, although certain aspects of this dynamical system 
can be described in terms of local field equations, the system as a whole 
is essentially nonlocal. There is solid physics behind this: the nonlocal
terms reflect the influence of scalar fluctuations within the past 
light-cone. Note that this is perfectly consistent with causality. Note 
further that, since $\ln(1-\eta/\eta')$ diverges logarithmically at $\eta'
= \eta$, the last integral cannot be recast in the form of an integral over 
a kernel ${\tt K} = {\tt K}(\eta,\eta')$ multiplied by the polarization 
vector, $\int_{\eta_0}^\eta{\tt K}(\eta,\eta') \epsilon_i(\vec k,\eta')$. 
Instead, its form is that of a nonlocal conductivity, $\int_{\eta_0}^\eta{
\tt K'}(\eta,\eta') \partial_0^{\prime} \epsilon_i(\vec k,\eta')$. We shall 
now address the relevance of these nonlocal (potentially dissipative) 
contributions.

\section{Late-time dynamics of superhorizon photons}
\label{Late-time dynamics of superhorizon photons}

\subsection{Analytical analysis}

We begin by performing a partial integration on Eq.~(\ref{eom:Ai:2}),
\begin{eqnarray}
&&(\partial_0^2 + k^2)\epsilon_i(\vec k,\eta)
 + a^2H^2 \,\frac{2\alpha}{\pi}
  \bigg\{
   \bigg(
       \ln\Big(\frac{k}{H}\Big)
     - \frac 54 + \gamma_E
   \bigg) \epsilon_i(\vec k,\eta)
\nonumber\\
 &&\hspace{4.5cm}
       -\; \int_{\eta_0}^\eta d\eta^\prime
           \Big[
                {\rm ci}(k\Delta\eta)+\frac{\sin(k\Delta\eta)}{k\eta^\prime}
           \Big]\partial_0{^\prime}\epsilon_i(\vec k,\eta^\prime)\bigg\}
\nonumber\\
&& \hspace{4.5cm} = \; a^2H^2 \,\frac{2\alpha}{\pi}
  \bigg[{\rm ci}(k\Delta\eta_0)+\frac{\sin(k\Delta\eta_0)}{k\eta_0}\bigg]
            \epsilon_i(\vec k,\eta_0)
\,.\qquad
\label{eom:IntDE:1}
\end{eqnarray}
It is now convenient to rewrite this equation in terms of the number of 
e-foldings, $N \equiv \ln(a) = - \ln(-H\eta)$,
\begin{eqnarray}
&&\!\!\!\!\!\!\!\!\!
(\partial_N^2 + \partial_N + w^2a^{-2})\widetilde\epsilon_i(w,N)
 + \frac{2\alpha}{\pi}
  \bigg\{
   \bigg(
         \ln\big(w\big)
       - \frac 54 + \gamma_E  
   \bigg)\widetilde\epsilon_i(w,N)
\nonumber\\
&& \hspace{4.7cm} 
       -\; \int_{0}^N dN^\prime
           \bigg[
                {\rm ci}\big(w[{a^\prime}^{-1}-{a}^{-1}]\big)
              -\frac{\sin\big(w[{a^\prime}^{-1}-{a}^{-1}]\big)}
                    {w{a^\prime}^{-1}}
           \bigg]\partial_0{^\prime}\widetilde\epsilon_i(w,N^\prime)\bigg\}
\nonumber\\
&&\hspace{4.7cm} = \;\frac{2\alpha}{\pi}\bigg[
                                            {\rm ci}\Big(w(1-1/a)\Big)
                                         -  \frac{\sin\Big(w(1-1/a)\Big)}{w}
                                      \bigg]\widetilde\epsilon_i(w,0)
\, .\qquad
\label{eom:IntDE:2}
\end{eqnarray}
The new dependent variable is, $\widetilde\epsilon_i(w,N) \equiv 
\epsilon_i(k,\eta)$, where $w \equiv k/H\gg 1$. The scale factors are $a =
{\rm e}^N$ and $a^\prime ={\rm e}^{N^\prime}$. When $N$ becomes large the 
terms containing negative powers of $a$ can be dropped to give,
\begin{eqnarray}
&& \!\!\!\!\!\!\!\!
(\partial_N^2 + \partial_N )\widetilde\epsilon_i(w,N)
 + \frac{2\alpha}{\pi}
  \bigg\{
   \bigg(
         \ln\big(w\big)
       - \frac 54 + \gamma_E  
   \bigg) \widetilde\epsilon_i(w,N)
\nonumber\\
&& \hspace{1.2cm} 
       - \int_{0}^N dN^\prime
           \bigg[
                {\rm ci}\big(w{a^\prime}^{-1}\big)
              -\frac{\sin\big(w{a^\prime}^{-1}\big)}
                    {w{a^\prime}^{-1}}
           \bigg]\partial_{N'}\widetilde\epsilon_i(w,N^\prime)\bigg\}
        \simeq  \;\frac{2\alpha}{\pi}\bigg[
                                            {\rm ci}(w)
                                         -  \frac{\sin(w)}{w}
                                      \bigg]\widetilde\epsilon_i(w,0)
\,.\qquad
\label{eom:IntDE:3}
\end{eqnarray}

It is easy to see from equation (\ref{eom:IntDE:3}) that 
$\widetilde\epsilon_i(w,N)$ approaches a constant plus an exponentially small
term. First note that the assumption is self consistent. If we make it then, 
for large enough $N'$, the integrand of the nonlocal term falls off. This 
means that the integral is independent of its upper limit and the constant 
is determined by the relation,
\begin{equation}
\bigg(\!\! \ln\big(w\big) - \frac 54 + \gamma_E \!\!\bigg) \widetilde{
\epsilon}_i(w,\infty) - \! \int_{0}^{\infty} \!\!\!\! dN^\prime \bigg[ 
{\rm ci}\big(w{a^\prime}^{-1} \big) -\frac{\sin\big(w{a^\prime}^{-1}\big)}{
w{a^\prime}^{-1}} \bigg] \partial_{N'} \widetilde\epsilon_i(w,N^\prime) 
\simeq \! \bigg[ {\rm ci}(w) - \frac{\sin(w)}{w} \bigg] \widetilde{
\epsilon}_i(w,0) \, .
\end{equation}
However, because the integral depends upon $\partial_{N'} \widetilde{
\epsilon}_i(w,N')$ for finite $N'$, when it is still significant, the actual
value of $\widetilde{\epsilon}_i(w,\infty)$ must be determined numerically.

To get the rate at which $\widetilde{\epsilon}_i(w,N)$ approaches 
$\widetilde{\epsilon}_i(w,\infty)$ we differentiate (\ref{eom:IntDE:3}) with 
respect to $N$. Neglecting exponentially small terms gives the following 
local equation for $\partial_N \widetilde{\epsilon}_i(w,N)$,
\begin{eqnarray}
\Big\{\partial_N^2 + \partial_N
 + \frac{2\alpha}{\pi}
\Big(N - \frac 14\Big)\Big\}\partial_N\widetilde\epsilon_i(w,N)
           \simeq  0
\, . \qquad
\label{eom:IntDE:5}
\end{eqnarray}
The general solution can be expressed in terms of Airy functions,
\begin{equation}
 \partial_N\widetilde\epsilon_i(w,N) 
        \simeq a^{-1/2}\Bigl\{C_A\, {\rm Ai}\Big(-(2\alpha/\pi)^{1/3}
                                     \Bigl[
                                           N - \frac 14 - \frac{\pi}{8\alpha}
                                     \Bigr]
                                 \Big) 
 \,+\, C_B\, {\rm Bi}\Big(-(2\alpha/\pi)^{1/3}
                                     \Bigl[
                                           N - \frac 14 - \frac{\pi}{8\alpha}
                                     \Bigr]
                                 \Big) 
                        \Bigr\}
\, . \; \label{Airy solution:2}
\end{equation}
Making use of the asymptotic expansion of the Airy functions~\footnote{The 
asymptotic expansion for the Airy functions, 
$$
  {\rm Ai}(-z) \sim \frac{1}{\pi^{1/2}z^{1/4}}
                    \Big[
                         \cos\Big(\frac 23 z^{3/2} - \frac{\pi}{4}\Big)
                      +  O(z^{-3/2})
                    \Big]\,,\qquad |z|\gg 1
$$
$$
  {\rm Bi}(-z) \sim \frac{1}{\pi^{1/2}z^{1/4}}
                    \Big[
                         \sin\Big(\frac 23 z^{3/2} - \frac{\pi}{4}\Big)
                      +  O(z^{-3/2})
                    \Big]\,,\qquad |z|\gg 1
\,.
\label{Airy functions:aasymptotic}
$$
}, 
Eq.~(\ref{Airy solution:2}) can be integrated to give asymptotically, 
\begin{eqnarray}
\!\! \widetilde \epsilon_i(w,N) = \widetilde \epsilon_i(w,\infty) 
         +  {\rm e}^{-N/2}
  \Bigg\{\! \frac{C_A\sin\Big(\varphi[N,\alpha]\Big)
            -  C_B\cos\Big(\varphi[N,\alpha]\Big)}
                 {\pi^{1/2}(2\alpha/\pi)^{7/12}
                        \Big(N-\frac 14 - \frac{\pi}{8\alpha}\Big)^{3/4}}
    + O(N^{-\frac 54})
  \! \Bigg\}
\,, \;\;  N\gg \ln(w)\,, 
 \quad
\label{Airy solution:epsiloni}
\end{eqnarray}
where,
\begin{equation}
\varphi[N,\alpha] \equiv \frac 23\Big[\frac{2\alpha}{\pi}\Big]^{1/2}
                        \Big[N-\frac 14 - \frac{\pi}{8\alpha}\Big]^{3/2}
                            - \frac\pi 4 \, .
\label{varphi:def}
\end{equation}
Of course the constants $C_A$ and $C_B$ are fixed by the behavior of 
$\widetilde{\epsilon}_i(w,N)$ before the asymptotic form obtains. They
can only be determined numerically.

\subsection{Numerical analysis}

By performing a numerical study of Eq.~(\ref{eom:Ai:2}),
we now confirm the analytical results 
of section~\ref{Late-time dynamics of superhorizon photons} above.
\begin{figure}[tb]
\vskip 0.1in
\leftline{\epsfig{file=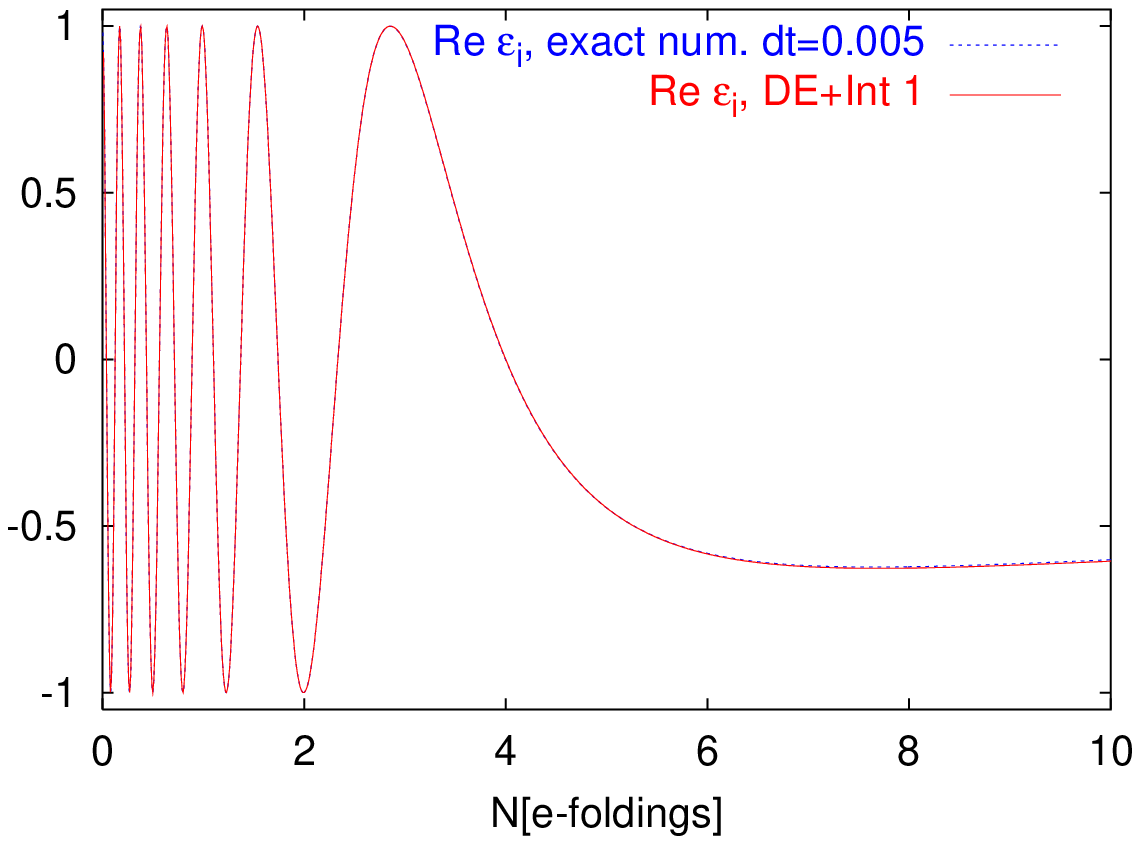,width=3.3in,height=3.2in }}
\vskip -3.2in
\rightline{\epsfig{file=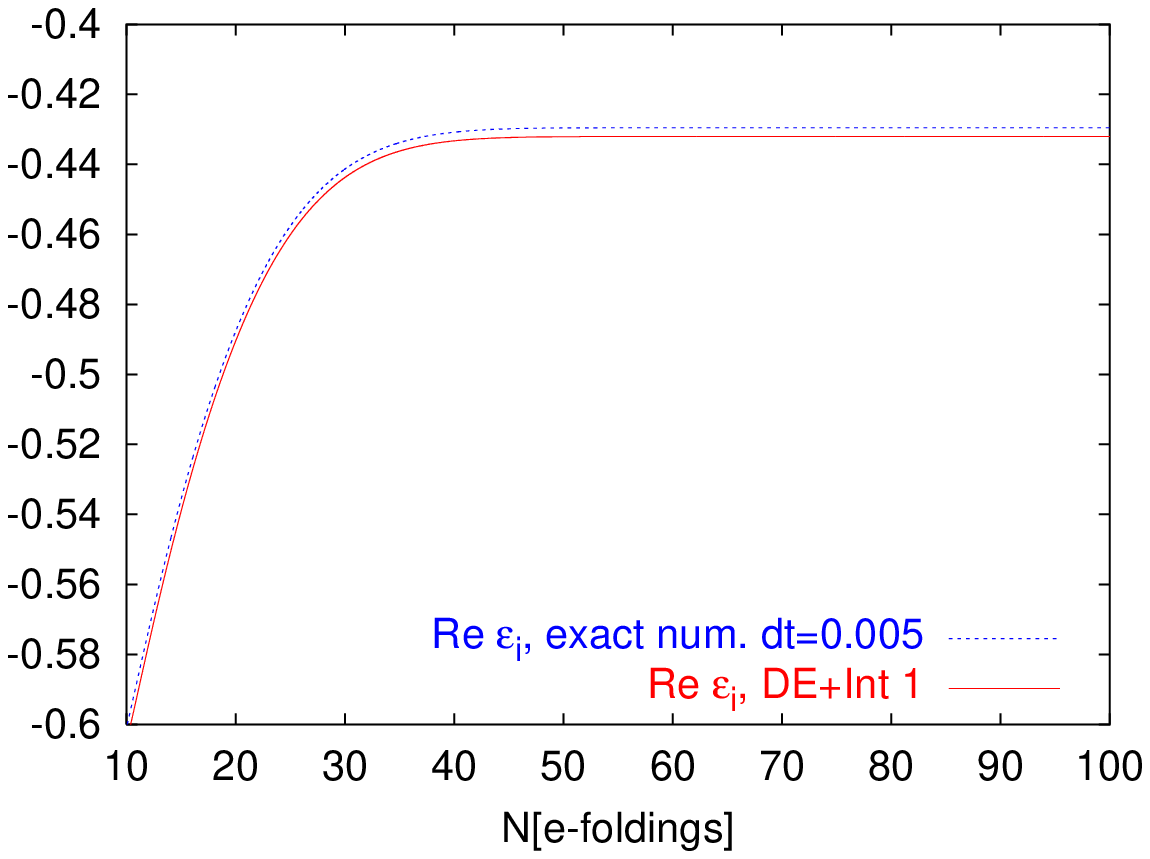, width=3.3in,height=3.2in }
   \hskip 0.in}
\vskip -0.1in
\caption{
A comparison of the exact (numerical) solution of  
the integro-differential equation~(\ref{eom:Ai:2}) with that of 
Eqs.~(\ref{DE:zeta}-\ref{eom:DE+Int1}) for $k = 40 H$.
At late times the photon amplitude, $\epsilon_i$, approaches a constant.
The difference between the two solutions is smaller or of the order
the thickness of the lines, and quantitatively, it reaches about 
0.7\%.
\label{figure-I} 
        }
\end{figure}
The following approximation turns out to be quantitatively justified.
Upon neglecting the latter two integrals, Eq.~(\ref{eom:Ai:2})
reduces to,
\begin{eqnarray}
&&(\partial_0^2 + k^2)\epsilon_i(\vec k,\eta)
 + a^2 \,\frac{2\alpha H^2}{\pi}
   \Big\{\Big[\ln(a)
   -\gamma_E + \ln\Big(\frac{H}{2\lambda}\Big)\Big]
            \epsilon_i(\vec k,\eta)
       - \zeta(\vec{k},\eta)
   \Big\}
 \simeq 0
\,,\qquad
\label{eom:DE+Int1}
\end{eqnarray}
where we define, 
\begin{equation}
  \zeta(\vec{k},\eta) \equiv \frac{H^2}{k} \int_{\eta_0}^\eta d\eta' {a'}^2
            \sin(k\Delta\eta)\epsilon_i(\vec k,\eta')
\,,
\label{zeta}
\end{equation}
The function $\zeta(\vec{k},\eta)$ obeys the following differential equation 
and initial conditions,
\begin{equation}
   \Big(\frac{\partial^2}{\partial \eta^2} + k^2\Big) \zeta = a^2 H^2 
\epsilon_i \qquad \zeta(\vec{k},\eta_0)=0
,\quad \frac{\partial}{\partial \eta}\zeta(\vec{k},\eta)|_{\eta_0}=0
\,.
\label{DE:zeta}
\end{equation}

To study how quantitative this approximation is, in figure~\ref{figure-I}
we compare the exact (numerical) solution of Eq.~(\ref{eom:Ai:2})
with the approximate one obtained by solving
Eqs.~(\ref{eom:DE+Int1}--\ref{DE:zeta}). 
As can be seen in figure~\ref{figure-I}, the agreement is quite good. This
implies that the two final integrals in~(\ref{eom:Ai:2}) are dynamically 
irrelevant.
  
\begin{figure}[tbp]
\vskip -0.25in
\centerline{\epsfig{file=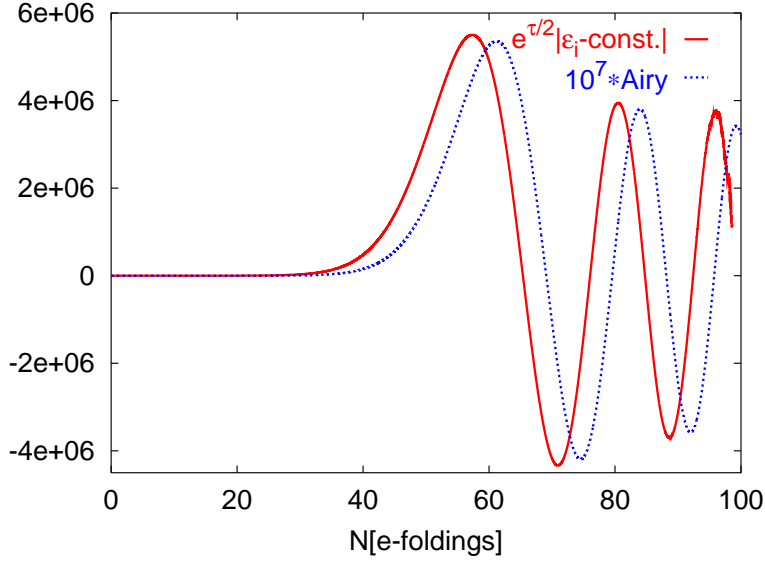,width=4.3in}}
\caption{A comparison of the solution of~(\ref{eom:DE+Int1}--\ref{DE:zeta})
with that of the Airy oscillator~(\ref{eom:IntDE:5}).
        }
\label{figure-III} 
\vskip -0.1in
\end{figure}

Figure~\ref{figure-I} shows that, at asymptotically late times, 
$\widetilde \epsilon_i(w,N)$ is dominated by the constant value $\widetilde 
\epsilon_i(w,\infty)$, whose real part is slightly suppressed with respect 
to its classical asymptotic value of $\cos(40) \approx -.67$.
In figure~\ref{figure-III} we subtract the asymptotic constant 
from the solution of Eqs.~(\ref{eom:DE+Int1}-\ref{DE:zeta}), multiply by
$e^{N/2}$, and compare the result with the Airy function. The figure 
displays an oscillatory, slightly damped profile, in reasonably good 
agreement with our result (\ref{Airy solution:epsiloni}).

\section{Discussion}
\label{Discussion}

It is illuminating to translate our results for the vector potential,
$A_i(x) = \epsilon_i(\vec{k},\eta) e^{i \vec{k} \cdot \vec{x}}$, into 
statements about the behavior of the electric and magnetic fields. In
Coulomb gauge and using conformal coordinates, the densities of electric 
and magnetic field lines per physical 3-volume are given by the following 
expressions,
\begin{equation}
E^i(x) = \frac{\partial_0 A_i(x)}{a^2(\eta)} \qquad , \qquad B^i(x) =
-\frac{\epsilon^{ijk} \partial_j A_k(x)}{a^2(\eta)} \; .
\end{equation}
For spatial plane waves with $a(\eta) = e^N$ and $w = k/H$ we have,
\begin{equation}
e^{-i \vec{k} \cdot \vec{x}} E^i(x) = H e^{-N} \partial_N \widetilde{
\epsilon}_i(w,N) \qquad , \qquad e^{-i \vec{k} \cdot \vec{x}} B^i(x) = 
-i \epsilon^{ijk} k^j e^{-2N} \widetilde{\epsilon}_k(w,N) \; .
\end{equation}
Keeping track only of powers of the scale factor, we can summarize the
results of section \ref{Late-time dynamics of superhorizon photons} by
saying that the full quantum system results in the following asymptotic
behavior,
\begin{equation}
\eta^{\mu\nu} \eta^{\rho\sigma} \partial_{\rho} F_{\sigma\nu}(x) \!
+ \!\! \int \! d^4x' [\mbox{}^{\mu} \Pi_{\rm ret}^{\nu}]\!(x,x') A_{\nu}(x') 
\! = \! 0 \quad \Longrightarrow \quad E^i \sim a^{-\frac32} \quad {\rm and} 
\quad B^i \sim a^{-2} \; . \label{oursys}
\end{equation}

It is interesting to contrast the asymptotic behavior (\ref{oursys}) of 
the full, nonlocal system with that of various local models. The tree order 
system gives,
\begin{equation}
\eta^{\mu\nu} \eta^{\rho\sigma} \partial_{\rho} F_{\sigma\nu} \! = \! 0
\quad \Longrightarrow \quad E^i \sim a^{-2} \quad {\rm and} \quad B^i 
\sim a^{-2} \; . \label{tree}
\end{equation}
Although the behavior of the magnetic fields agree as regards powers of $a$,
we saw in Figure~\ref{figure-I} that the magnetic field of the actual 
system analyzed in section III B is smaller by a factor of about 2/3. On the 
other hand, the electric field of the actual system is enhanced by an 
enormous factor of $a^{\frac12}$.

Adding a Proca term with fixed mass $m_{\gamma} = g H$ results in the 
following asymptotic behavior,
\begin{equation}
\eta^{\mu\nu} \Bigl(\eta^{\rho\sigma} \partial_{\rho} F_{\sigma\nu} -
g^2 H^2 a^2 A_{\nu} \Bigr) \! = \! 0 \quad \Longrightarrow \quad E^i 
\sim a^{-\frac32 + \frac12 \sqrt{1 - 4 g^2}} \quad {\rm and} \quad B^i 
\sim a^{-\frac52 + \frac12 \sqrt{1 - 4 g^2}} \; . \label{fixedm}
\end{equation}
The best perturbative fit to the actual system has $g^2 = 2 \alpha\ln(w)/\pi$
\cite{ProkopecTornkvistWoodard:2002a,ProkopecTornkvistWoodard:2002b} but
we see that the agreement between (\ref{oursys}) and (\ref{fixedm}) is
not good. The constant mass system has too much electric field and not
enough magnetic field.

If the mass term grows as $m^2_{\gamma} = g^2 H^2 \ln(a)$ one finds,
\begin{equation}
\eta^{\mu\nu} \Bigl(\eta^{\rho\sigma} \partial_{\rho} F_{\sigma\nu} -
g^2 H^2 \ln(a) a^2 A_{\nu} \Bigr) \! = \! 0 \quad \Longrightarrow \quad E^i 
\sim a^{-\frac32} \quad {\rm and} \quad B^i \sim a^{-\frac52} \; . 
\label{mgrows}
\end{equation}
This model is suggested by mean field theory 
\cite{DavisDimopoulosProkopecTornkvist:2000,
TornkvistDavisDimopoulosProkopec:2000,DimopoulosProkopecTornkvistDavis:2001}
with $g^2 = 2 \alpha/\pi$. It gives good agreement for the electric field 
but results in too little magnetic field. In particular note that, whereas 
the actual polarization vector approaches a nonzero constant, $\widetilde{
\epsilon}_i(w,\infty)$, the polarization vector of (\ref{mgrows}) approaches 
zero.

While none of the local models described above gives a perfect fit, the
effect of vacuum polarization is to vastly enhance the asymptotic electric 
field with respect to its classical value in a manner most nearly described
by the growing mass of (\ref{mgrows}). Without regard to local analogues,
it is very apparent that vacuum polarization alters the kinematical
properties of superhorizon photons during inflation. On the other hand, the
fact that the asymptotic polarization vector is somewhat suppressed from
its classical value seems to indicate that there is no significant production 
of photons.

Exactly the opposite results seem to pertain for massless fermions which are
Yukawa coupled to a massless, minimally coupled 
scalar~\cite{ProkopecWoodard:2003b}. In that case chirality conservation
protects the particle from gaining a mass, but the amplitude of the wave
function grows faster than exponentially during inflation. The physical 
explanation seems to be copious particle production.

\begin{acknowledgments}

 It is a pleasure to acknowledge Ola T\"ornkvist's collaboration
in much of the work reported here. We have also profited from
conversations with Dietrich B\"odeker, Anne-Christine Davis, Konstantinos
Dimopoulos, Sasha Dolgov,  Hendrik J. Monkhorst, Glenn Starkman
and Tanmay Vachaspati. This work was partially supported by DOE contract 
DE-FG02-97ER41029 and by the Institute for Fundamental Theory of
the University of Florida.

\end{acknowledgments}

\eject

\section*{Appendix: Reduction of the polarization integral}
\label{Appendix: Reduction of the polarization integral}

 Here we show some details of the calculation of the inflationary  
contribution to the vacuum polarization integral in~(\ref{genMax})
with the general plane wave (\ref{vector:field:plane:wave})
for the photon field, 
$A_{\nu}(x') = \epsilon_\nu(\vec k,\eta')\; {\rm e}^{i\vec k\cdot \vec x'}$
$\;(k_i\epsilon_i = 0 = \epsilon_0)$. 

We begin our analysis by noting a useful identity, 
\begin{eqnarray}
  - [{}^\mu\!P^\nu]\; aa' 
  &=& -\, aa' \delta^\mu_0\delta^\nu_0 \nabla^2
      + \bar\eta^{\mu\nu}[aa' (\nabla^2-\partial_0^2)
                         + H^2 a^2{a'}^2(1-\Delta\eta\partial_0)
                          ]
\nonumber\\
     && -\, Ha{a'}^2 \delta_\mu^0\bar\partial^\nu
      + Ha^2a' \bar\partial^\mu\delta^\nu_0
      + aa'(\bar\partial^\mu\delta^\nu_0 
                   + \delta^\mu_0\bar\partial^\nu)\partial_0
      - aa' \bar\partial^\mu\bar\partial^\nu
\, .
\label{A:transverse-on:aa'}
\end{eqnarray}
In deriving this we made use of the relations, 
\begin{equation}
 \partial_\rho^{\,\prime} aa' = aa'\partial_\rho^{\,\prime} 
                          + Ha{a'}^2 \delta_\rho^0
\qquad {\rm and} \qquad
 \partial_\sigma aa' = aa'\partial_\sigma
                          + Ha^2a' \delta_\sigma^0
\,,
\label{A:partial:aa'}
\end{equation}
which imply, 
\begin{equation}
 \partial_\rho^{\,\prime}\partial_\sigma aa' 
     = H^2a^2{a'}^2 \delta_\rho^0\delta_\sigma^0
     + Ha{a'}^2\delta_\rho^0\partial_\sigma
     - Ha^2a'\partial^{\,\prime}_\rho\delta_\sigma^0
     - aa'\partial^{\,\prime}_\rho\partial_\sigma
\, .
\label{A:partialpartial:aa'}
\end{equation}
Note also that we employ the notation,
\begin{equation}
 \bar\partial^\mu \equiv \eta^{\mu i}\partial_i
\,,\qquad \bar \eta^{\mu\nu} \equiv \eta^{\mu\nu} + \delta^\mu_0\delta^\nu_0
\,.
\label{A:notation:eta+partial}
\end{equation}
When a derivative operator stands on the right of $aa'$, then it acts on a
function of $x^{\mu} - x^{\prime \mu}$ and we can write $\partial_\rho^{\,
\prime} = - \partial_\rho$. Moreover, due to the (spatial) transversality 
of the photon field, $\bar\partial^\nu A_\nu(x) = 0$, the terms 
in~(\ref{A:transverse-on:aa'}) containing $\bar\partial^\nu$ vanish 
identically. 

Consider first the $0$-th component of the photon field
equation~(\ref{genMax}),
\begin{equation}
 - \nabla^2 A_0 - \frac{\alpha H^2}{8\pi^2}a \nabla^2 \partial^2 
   \int d^4 x' \theta(\Delta\eta)\theta(\Delta \tau^2)
   \Bigl[
         \ln({H^\prime}^{\,2}\Delta\tau^2) + 1
   \Bigr] a' A_0(x') = 0
\, ,
\label{A:A0equation}
\end{equation}
where we made use of 
Eqs.~(\ref{vacuum:pol:retarded:2}-\ref{vacuum:pol:retarded:3})
and~(\ref{A:transverse-on:aa'}), and of 
the Coulomb gauge condition, $\nabla\cdot \vec A = 0$. 
Note that only the first term in~(\ref{A:transverse-on:aa'}) contributes
to~(\ref{A:A0equation}). Furthermore, due to the spatial transverse
structure of the Lorentz breaking term, $[{}^\mu\!\bar P^\nu]\delta n^2$,
it does not contribute to~(\ref{A:A0equation}). 

At $\eta = \eta_0$ the integral drops out of (\ref{A:A0equation}) and we
conclude that $\nabla^2 A_0(\vec{x},\eta_0) = 0$. We can make use of 
residual gauge freedom to make the unique solution be $A_0(\vec{x},\eta_0)
= 0$. The same thing can be done for the first conformal time derivative.
But then the equation implies that $A_0(\vec{x},\eta)$ vanishes for all
$\eta$. Therefore, it is consistent to assume $\epsilon_0(\vec k,\eta) = 0$.
This discussion also implies that the terms in~(\ref{A:transverse-on:aa'}) 
which contain $\delta_\nu^0$ cannot contribute to the photon dynamical 
equation~(\ref{genMax}).

Consider now the spatial components of equation~(\ref{genMax}).
We start with the de Sitter contribution to the electric susceptibility, 
$\chi_e|_{\rm de\; Sitter}$~(\ref{vacuum:pol:retarded:2}),
which with the help of~(\ref{A:transverse-on:aa'})
(only the terms proportional to $\bar\eta^{\mu\nu}$ contribute),
\begin{eqnarray}
&&-\eta_{i\mu}
      \int\! d^4x' [{}^\mu\!P^\nu] \chi_e(x,x')|_{\rm de\;Sitter}\;A_{\nu}(x')
\;=\; \frac{\alpha H^2}{2\pi k^3}a {\rm e}^{i\vec k\cdot\vec x}
       (\partial_0^2 + {k}^{\,2})^2\int_{\eta_0}^\eta d\eta' a'
         \Xi_1\Big(k\Delta\eta,\frac{k}{H^\prime}\Big)\epsilon_i(\vec k,\eta')
\nonumber\\
 &&\hspace{4.1cm}
  -\, \frac{\alpha H^4}{2\pi k^3}a^2 {\rm e}^{i\vec k\cdot\vec x}
       (\partial_0^2 + {k}^{\,2})\!\int_{\eta_0}^\eta d\eta' 
       {a'}^2(1-\Delta\eta\partial_0)
         \Xi_1\Big(k\Delta\eta,\frac{k}{H^\prime}\Big)\epsilon_i(\vec k,\eta')
\, . \qquad
\label{A:chi_e:deSitter}
\end{eqnarray}
Here $\eta_0 = -1/H$ denotes the conformal time at the beginning of
inflation, at which $a(\eta_0) = 1$. We define $k \equiv \Vert \vec k \Vert$, 
$\Delta \eta = \eta - \eta '$. We also made use of the following elementary 
integrals, 
\begin{eqnarray}
&& \int_0^{\Delta\eta} d^3 \Delta x\, {\rm e}^{-i\vec k\cdot \Delta \vec x}
        \Big\{
              \ln\Big[{H^\prime}^{\,2}(\Delta\eta^2-\Vert\Delta\vec x\Vert^2)
                  \Big]
           +  1
        \Big\}
\nonumber\\
&&\hspace{4.cm}
=  4\pi \int_0^{\Delta\eta} r^2 dr \frac{\sin(kr)}{kr}
        \Big\{
              \ln\Big[{H^\prime}^{\,2}(\Delta\eta^2-\Vert\Delta\vec x\Vert^2)
              \Big]
           +  1
        \Big\}
\nonumber\\
&&\hspace{4.cm}
=\frac{4\pi}{k}\Delta\eta^2 \int_0^{1} x dx \sin(k\Delta\eta x)
        \Big\{
              \ln(1-x^2) + 2\ln({H^\prime}\Delta\eta)+1
        \Big\}
\nonumber\\
&&\hspace{4.cm}
=\frac{4\pi}{k^3}\Xi_1(k\Delta\eta,k/{H^\prime})
\,,
\quad
\end{eqnarray}
where $\Delta\vec x = \vec x - \vec x^{\,\prime}$, 
\begin{eqnarray}
  \Xi_n(z,\zeta) &\equiv & [\sin(z)-x\cos(z)]
                           [2\ln(z/\zeta)+n] + z^2\xi(z)
\nonumber\\
z^2\xi(z) &\equiv& z^2\int_0^1 z'dz' \sin(zz')\ln(1-{z'}^2)
\nonumber\\
          &=& 2\sin(z)
          - [\cos(z)+z\sin(z)][{\rm si}(2z)+\pi/2]
          + [\sin(z)-z\cos(z)][{\rm ci}(2z)-\gamma_E-\ln(z/2)]
\nonumber\\
{\rm si}(z) &\equiv & -\int_z^\infty\frac{\sin(t)}{t}dt
               =      \int_0^z\frac{\sin(t)}{t}dt - \frac{\pi}{2}
\nonumber\\
{\rm ci}(z) &\equiv & -\int_z^\infty\frac{\cos(t)}{t}dt
               =      \int_0^z\frac{\cos(t)-1}{t}dt + \gamma_E + \ln(z)
\,,
\end{eqnarray}
and $\gamma_E \equiv - \psi(1) = -(d/dz)\ln[\Gamma(z)]|_{z=1} = 
0.577\, 215\, 664.. $ is Euler's constant. 
Now making use of, 
\begin{eqnarray}
Y_1(z,\zeta)&\equiv& \frac 12 (\partial_z^2 + 1)\Xi_1(z,\zeta)
\label{A:Y1}
\\
     &=& \sin(z)\big[2\ln(2z/\zeta)+2
      +  {\rm ci}(2z)-\gamma_E-\ln(2z)\big] 
      -  \cos(z)\big[{\rm si}(2z)+\pi/2\big] 
\nonumber\\
\!\!\!\!
 (z\partial_z - 1)(\partial_z^2 + 1)\Xi_1(z,\zeta)
      &=& 4\sin(z) - 2\Xi_2(z,\zeta)
\,,
\end{eqnarray}
the integral~(\ref{A:chi_e:deSitter}) reduces to  
\begin{eqnarray}
&&\!\!\!\!\!\!
   -\eta_{i\mu}\int\! d^4x' [{}^\mu\!P^\nu] \chi_e(x,x')|_{\rm de\;Sitter}\;
                   A_{\nu}(x')
\;=\; \frac{\alpha H^2}{\pi k}a {\rm e}^{i\vec k\cdot\vec x}
       (\partial_0^2 + {k}^{\,2})\int_{\eta_0}^\eta d\eta' a'
           Y_1\Big(k\Delta\eta,\frac{k}{H^\prime}\Big)\epsilon_i(\vec k,\eta')
\nonumber\\
 &&\hspace{5.cm}
  +\, \frac{\alpha H^4}{\pi k}a^2 {\rm e}^{i\vec k\cdot\vec x}
       \int_{\eta_0}^\eta d\eta' {a'}^2
       \Big[
            2\sin(k\Delta\eta)-\Xi_2\Big(k\Delta\eta,\frac{k}{H^\prime}\Big)
       \Big]
       \epsilon_i(\vec k,\eta')
\,.
\qquad
\label{A:chi_e:deSitter:2}
\end{eqnarray}

The contribution from $\delta n^2$ to the spatial components 
of Eq.~(\ref{genMax}) can be obtained quite easily by the methods
analogous to those employed for the contribution from
$\chi_e|_{\rm de\; Sitter}$; the result can be written as,
\begin{eqnarray}
\eta_{i\mu}\int\! d^4x' [{}^\mu\!\bar P^\nu] \delta n^2(x,x')\; A_{\nu}(x')
\;=\; \frac{\alpha H^4}{\pi k}a^2 {\rm e}^{i\vec k\cdot\vec x}
         \int_{\eta_0}^\eta d\eta' {a'}^2
         \Xi_2\Big(k\Delta\eta,\frac{k}{H^\prime}\Big)\epsilon_i(\vec k,\eta')
\,.
\quad
\label{A:delta n2}
\end{eqnarray}
Remarkably, this contribution is fully canceled by the analogous 
one in~(\ref{A:chi_e:deSitter:2}). Keeping 
the terms that contribute at order $a^2$, we have, 
\begin{eqnarray}
\eta_{i\mu}\!\int\! d^4x' [{}^\mu\!\Pi_{\rm ret}^\nu](x,x')\, A_{\nu}(x')
&\!=\!&\eta_{i\mu}\!\int\! d^4x' 
                        \Big(\!
                          \!-\! [{}^\mu\!P^\nu] \chi_e(x,x')|_{\rm de\;Sitter}
                          \!+\!  [{}^\mu\!\bar P^\nu] \delta n^2(x,x')
                        \Big) A_{\nu}(x')
 \!+\! O(a)
\nonumber\\
 &\!=\!&\; \frac{\alpha H^2}{\pi k}a {\rm e}^{i\vec k\cdot\vec x}
       (\partial_0^2 + {k}^{\,2})\int_{\eta_0}^\eta d\eta' a'
           Y_1\Big(k\Delta\eta,\frac{k}{H^\prime}\Big)\epsilon_i(\vec k,\eta')
\nonumber\\
 &\!+\!&
  \frac{2\alpha H^4}{\pi k}a^2 {\rm e}^{i\vec k\cdot\vec x}
       \int_{\eta_0}^\eta d\eta' {a'}^2
       \sin(k\Delta\eta)\epsilon_i(\vec k,\eta')
 + O(a)
\,,
\label{A:chi_e:deSitter+delta n2}
\end{eqnarray}
where $O(a)$ indicates that the neglected terms can contribute
to the vacuum polarization at most linearly in $a$. 

 Now from the small argument expansions of the sine and cosine integral
({\it cf.} Eq.~[8.232] in Ref.~\cite{GradshteynRyzhik:1965}), 
\begin{eqnarray}
  {\rm si}(z) &=& -\frac{\pi}{2} + \sum_{n=1}^\infty
                 \frac{(-1)^{n+1}z^{2n-1}}{(2n-1)\,(2n-1)!}
\,,\qquad |z|\ll 1
\nonumber\\
  {\rm ci}(z) &=& \gamma_E + \ln(z) + \sum_{n=1}^\infty
                 \frac{(-1)^{n}z^{2n}}{(2n)\,(2n)!}
\,,\qquad\;\; |z|\ll 1
\label{si:ci:small z}
\end{eqnarray}
we obtain easily the following small argument expansion of $Y_1$, 
\begin{eqnarray}
  Y_1(z,\zeta) &=& 2z\ln\Big(\frac{2z}{\zeta}\Big) 
                -  \frac 13 z^3\Big[
                                    \ln\Big(\frac{2z}{\zeta}\Big) 
                                 - \frac{31}{12}
                               \Big]
                + O(z^5\ln(z),z^5)
\, .
\label{A:Y1:expansion}
\end{eqnarray}
This implies that, attempting to act with the derivative $\partial_0^2$ 
on the integral in~(\ref{A:chi_e:deSitter+delta n2}), would result  
in a divergent contribution at the upper limit of integration, $\eta'=\eta$. 
In order to overcome this difficulty, observe first
that one time derivative may be taken,
\begin{eqnarray}
\frac{\partial_0}{k} \int_{\eta_0}^\eta d\eta' a'
         Y_1\Big(k\Delta\eta,\frac{k}{H^\prime}\Big)\epsilon_i(\vec k,\eta')
   =  \int_{\eta_0}^\eta d\eta' a'
         \Big[\partial_z Y_1\Big(z,\frac{k}{H^\prime}\Big)\Big]_{z=k\Delta\eta}
         \epsilon_i(\vec k,\eta')
\, .
\end{eqnarray}
Note we can write,
\begin{eqnarray}
\partial_z Y_1\big(z,\zeta\big)
    &=& \cos(z)\big[2\ln(2z/\zeta)+2+{\rm ci}(2z)-\gamma_E-\ln(2z)\big]
     +  \sin(z)\big[{\rm si}(2z)+\pi/2\big] 
\nonumber\\
    &=& \Big[\partial_z Y_1\big(z,\zeta\big)-2\ln(2z/\zeta)-2\Big]
     +  2\ln(2z/\zeta) + 2
\,,
\end{eqnarray}
where in the last line we have added and subtracted the troublesome term
which is logarithmically divergent at the upper limit of integration,
$\eta'\rightarrow \eta$. We can now act with the second time derivative
on the first term, to arrive at,
\begin{eqnarray}
\frac{\partial^2_0+k^2}{k} \int_{\eta_0}^\eta\! d\eta' a'
     Y_1\Big(k\Delta\eta,\frac{k}{H^\prime}\Big)\epsilon_i(\vec k,\eta^\prime)
  &=& - 2\int_{\eta_0}^\eta\! d\eta' a'
            \frac{1-\cos(k\Delta\eta)}{\Delta\eta}
                            \epsilon_i(\vec k,\eta^\prime)
\nonumber\\
  &+& 2\partial_0\! \int_{\eta_0}^\eta
               d\eta' a'
            \Big[\ln(2H^\prime\Delta \eta)+1\Big]\epsilon_i(\vec k,\eta^\prime)
\,.
\qquad
\label{Y1:integral}
\end{eqnarray}
Now from,
\begin{eqnarray}
a'\ln(2H^\prime\Delta\eta) &=& -\frac{1}{H\eta'}\Big[\ln(-2H^\prime\eta')
                                           -  \sum_{n=1}^\infty\frac{1}{n}
                                              \Big(\frac{\eta}{\eta'}\Big)^n
                                         \Big]
\nonumber\\
   &=& - \frac{1}{H}\frac{d}{d\eta'}
        \Big[
            \frac 12 \ln^2(-2H^\prime\eta')
          +  \sum_{n=1}^\infty\frac{1}{n^2}\Big(\frac{\eta}{\eta'}\Big)^n\,
        \Big]
\,,
\nonumber
\end{eqnarray}
we can rewrite the second integral in~(\ref{Y1:integral}) as follows,  
\begin{eqnarray}
\int_{\eta_0}^\eta\! d\eta' a'\ln(2H^\prime\Delta\eta)
                              \epsilon_i(\vec k,\eta^\prime) 
 &=& -\frac{1}{H}\bigg\{
        \Big[
             \frac 12 \ln^2(-2H^\prime\eta')
          +  \sum_{n=1}^\infty\frac{1}{n^2}\Big(\frac{\eta}{\eta'}\Big)^n\,
        \Big]\epsilon_i(\vec k,\eta')
                 \bigg\}\Big\vert_{\eta'=\eta_0}^{\eta'=\eta}
\nonumber\\
 &+& \int_{\eta_0}^\eta\! d\eta'
        \Big[
             \frac 12 \ln^2(-2H^\prime\eta')
          +  \sum_{n=1}^\infty\frac{1}{n^2}\Big(\frac{\eta}{\eta'}\Big)^n\,
        \Big]
            \frac{d}{d\eta'}\epsilon_i(\vec k,\eta^\prime)
\, .
\label{Y1:integral:3}
\end{eqnarray}
Its time derivative equals, 
\begin{eqnarray}
\partial_0\int_{\eta_0}^\eta\! d\eta' a'
                 \Big[
                      \ln(2H^\prime\Delta\eta)+1
                 \Big]
                 \epsilon_i(\vec k,\eta^\prime)
 &=& a\Big[\ln(-2H^\prime\eta)+1\Big]\epsilon_i(\vec k,\eta)
  +  a\ln\Big(1-\frac{\eta}{\eta_0}\Big)\epsilon_i(\vec k,\eta_0)
\nonumber\\
 &+& a\int_{\eta_0}^\eta\! d\eta' \ln\Big(1-\frac{\eta}{\eta'}\Big)
                            \frac{d}{d\eta'}\epsilon_i(\vec k,\eta^\prime)
\,.
\label{Y1:integral:4}
\end{eqnarray}
Upon collecting all of the terms, Eq.~(\ref{A:chi_e:deSitter+delta n2})
reduces to, 
\begin{eqnarray}
\eta_{i\mu}\!\int\! d^4x' [{}^\mu\!\Pi_{\rm ret}^\nu](x,x')\, A_{\nu}(x')
 &=& -\, a^2\frac{2\alpha H^2}{\pi}{\rm e}^{i\vec k\cdot\vec x}
    \bigg[-\Big(\ln(-2H^\prime\eta)+1\Big)\epsilon_i(\vec k,\eta)
\nonumber\\
 &&-\;
     \frac{H^2}{k}
     \int_{\eta_0}^\eta d\eta' {a'}^2
     \sin(k\Delta\eta)\epsilon_i(\vec k,\eta')
\nonumber\\
 &&+\; \frac 1a  \int_{\eta_0}^\eta\! d\eta' a'
            \frac{1-\cos(k\Delta\eta)}{\Delta\eta}
               \epsilon_i(\vec k,\eta^\prime)
\nonumber\\
 &&-\;
    \int_{\eta_0}^\eta\! d\eta'
     \ln\Big(1-\frac{\eta}{\eta'}\Big)
     \frac{d}{d\eta'}\epsilon_i(\vec k,\eta^\prime)
\nonumber\\
 &&-\; \ln\Big(1-\frac{1}{a}\Big)\epsilon_i(\vec k,\eta_0)
  \bigg]
\,,
\label{A:chi_e:deSitter+delta n2:2}
\end{eqnarray}
where we took account of $\eta_0 = -1/H$. When $a\gg 1$, 
the last local term, which depends on the initial photon amplitude
$\epsilon_i(\vec k,\eta_0)$, contributes as $O(a^{-1})$, such that,
in the limit when $a\gg 1$, it can be consistently neglected.

\end{document}